\pdfoutput=1

\documentclass[11pt,twoside,a4paper,cmspaper,final,collab]{cms-tdr}
\def\svnVersion{166532}\def\cmsCernNo{2012-049}\def\cmsCernDate{\today}\def\cmsMessage{Submitted to the Journal of High Energy Physics}

\begin{document}\cmsNoteHeader{EWK-11-012}

\hyphenation{had-ron-i-za-tion}
\hyphenation{cal-or-i-me-ter}
\hyphenation{de-vices}

\RCS$Revision: 114713 $
\RCS$HeadURL: svn+ssh://svn.cern.ch/reps/tdr2/papers/EWK-11-012/trunk/EWK-11-012.tex $
\RCS$Id: EWK-11-012.tex 114713 2012-04-07 14:20:23Z alverson $
\newcommand{\ZGS}{\ensuremath{\cmsSymbolFace{Z}/\gamma^*}\xspace} 
\newcommand{\ZGSB}{\ensuremath{\cmsSymbolFace{Z}/\gamma^*}+\,b-jet\xspace} 
\newcommand{\theory}{\ensuremath{(\mathrm{theory})}\xspace}
\cmsNoteHeader{EWK-11-012} 
\title{Measurement of the Z$/\gamma^*$+\,b-jet cross section in pp collisions at $\sqrt{s}=7$\,TeV}

\date{\today}

\abstract{

  The production of b jets in association with a Z$/\gamma^*$ boson is
  studied using proton-proton collisions delivered by the LHC at a
  centre-of-mass energy of 7\,TeV and recorded by the CMS
  detector. The inclusive cross section for Z$/\gamma^*$+\,b-jet
  production is measured in a sample corresponding to an integrated
  luminosity of 2.2\,fb$^{-1}$. The Z$/\gamma^*$+\,b-jet cross section
  with Z$/\gamma^*\rightarrow \ell\ell$ (where $\ell\ell =
  \mathrm{ee}$ or $\mu\mu$) for events with the invariant mass $60 <
  M_{\ell\ell} < 120$\,GeV, at least one b jet at the hadron level
  with $p_{\mathrm{T}} > 25$\,GeV and $|\eta|<2.1$, and a separation
  between the leptons and the jets of $\Delta R > 0.5$ is found to be
  $5.84 \pm 0.08\,(\mathrm{stat.}) \pm 0.72 \,(\mathrm{syst.})
  _{-0.55}^{+0.25} \,(\mathrm{theory})$\,pb. The kinematic properties
  of the events are also studied and found to be in agreement with the
  predictions made by the \textsc{MadGraph} event generator with the
  parton shower and the hadronisation performed by \textsc{pythia}.

}

\hypersetup{%
pdfauthor={CMS Collaboration},%
pdftitle={Measurement of the Z/gamma*+b-jet cross section in pp collisions at 7 TeV},%
pdfsubject={CMS},%
pdfkeywords={CMS, physics, electroweak, Z-boson, b-jets}}

\maketitle 

At the Large Hadron Collider (LHC), the measurement of the production
of a \ZGS boson in association with b quarks is important, both as a
benchmark channel to the production of the Higgs boson in association
with b quarks, and as a standard model background for searches for the
Higgs boson and new physics in final states with leptons and b jets.
The dominant contribution to \ZGSB production in proton-proton (pp)
collisions at the LHC centre-of-mass energy of 7\TeV comes from the
gluon-gluon interaction. Calculations of the cross section, driven by
perturbative QCD, are currently derived in two schemes: fixed-flavour
and variable-flavour. The fixed-flavour scheme allows only u, d, s,
and c quarks and gluons to participate in the hard scattering process,
with the b quarks produced explicitly in pairs from gluon
splitting. Complete calculations in this scheme with massive b quarks
at next-to-leading order (NLO) have recently been
released~\cite{Frederix:2011qg}. The variable-flavour scheme instead
allows the b quark to participate directly in the hard scattering by
integrating the gluon splitting process into the parton distribution
functions (PDFs). In this case NLO calculations have been performed
using massless b
quarks~\cite{Campbell:2003dd,Maltoni:2005wd,Campbell:2005zv}. To all
orders in perturbation theory, both schemes can be made exactly
identical. Still, at any finite order the results might differ
significantly, depending on the ordering of the perturbative
expansion.

This paper describes the measurement with the Compact Muon Solenoid
(CMS) detector of the pp\,$\rightarrow$\,\ZGSB cross section with
Z$/\gamma^*\rightarrow \ell\ell$ (where $\ell\ell = \Pe\Pe$ or
$\mu\mu$), including at least one b jet and any additional light jets,
and compares it with theoretical predictions in the variable-flavour
scheme. Distributions of the kinematic variables for jets and leptons
are also compared to Monte Carlo (MC) simulations using the tree-level
matrix-element calculations of \MADGRAPH (version
5.1.1.0)~\cite{MADGRAPH5} in the variable-flavour scheme with massless
b quarks, and using the \PYTHIA (version 6.424)~\cite{PYTHIA}
description of the parton shower and hadronisation processes with the
Z2 tune~\cite{Field:2010bc,QCD-10-010}. Multiparton interactions
(MPIs) are included in the \PYTHIA simulation. For systematic studies,
signal events are also simulated using the \SHERPA (version
1.3.0)~\cite{Gleisberg:2008ta} MC generator, in which the process is
computed at the leading order (LO) in the variable-flavour scheme, and
with the NLO fixed-flavour calculation of
{\footnotesize{a}}\MCATNLO~\cite{Frederix:2011qg} matched to the
\HERWIG (version 6.520)~\cite{Marchesini:1991ch,Corcella:2000bw} event
generator for the parton shower and the hadronisation part. Similar
measurements of the \ZGSB cross section were performed by the ATLAS
Collaboration~\cite{Aad:2011jn} at the LHC and by the
CDF~\cite{Aaltonen:2008mt} and D0~\cite{Abazov:2010ix} Collaborations
at the Tevatron $\mathrm{p}\mathrm{\bar{p}}$ collider. It should be
noted that in the latter cases the dominant contribution comes from
the quark-antiquark interaction.

The main backgrounds arise from the production of \ZGS with jets of
other flavours misidentified as b jets and from \ttbar+\,jets
events. Other processes such as QCD multijets, W\,+\,jets, single top,
and dibosons (WW, WZ) producing a final state with misidentified
leptons or b jets are found to give a negligible
contribution. Irreducible backgrounds such as ZZ and the associated
production of W and top, resulting in a final state with two genuine
leptons and a b jet, are also found to give a negligible contribution.

The CMS experiment uses a right-handed coordinate system, with the
origin at the nominal interaction point, the $x$ axis pointing to the
centre of the LHC ring, the $y$ axis pointing up (perpendicular to the
plane of the LHC ring), and the $z$ axis pointing along the
anticlockwise beam direction. The polar angle $\theta$ is measured
from the positive $z$ axis and the azimuthal angle $\phi$ is measured
in the $xy$ plane in radians. The pseudorapidity is defined as $\eta =
-\ln(\tan \frac{\theta}{2})$. The CMS detector features pixel and
silicon-strip trackers with coverage up to $|\eta|=2.4$ that, together
with a 3.8\,T solenoid magnet, allow for tracks with transverse
momentum (\pt) as low as 100\MeV to be reconstructed, and give a \pt
resolution of 1\% at 100\GeV. Also within the magnetic field are an
electromagnetic crystal calorimeter (ECAL) extending up to $|\eta| =
3.0$ with an electromagnetic transverse energy ($E_\mathrm{T}$)
resolution of about $3\%/\sqrt{E_\mathrm{T}/\!\GeV}$, and a hermetic
hadron calorimeter (HCAL) extending up to $|\eta| = 5.2$ with a
transverse hadronic energy resolution of
$~100\%/\sqrt{E_\mathrm{T}/\!\GeV}$. Embedded in the steel magnetic
field return yoke is an efficient muon system capable of
reconstructing and identifying muons up to $|\eta| = 2.4$. Further
details of the CMS detector may be found in Ref.~\cite{CMSExperiment}.

The data used in this analysis were collected between March and August
2011 and correspond to an integrated luminosity of $\mathcal{L} = 2.22
\pm 0.05$\fbinv~\cite{CMSLumi12}. The peak instantaneous luminosity
varied during this period from $1.0 \times 10^{32}$ to $2.5\times
10^{33}$\,cm$^{-2}$s$^{-1}$. The average number of inelastic
collisions per bunch crossing was 6.2 with an RMS of 2.9. The
selection of events with a \ZGS boson decaying to a pair of electrons
or muons is based on the selection used in the measurement of the
inclusive \ZGS cross section~\cite{CMS_paper_WZ}. Events are selected
using dielectron and dimuon triggers. The dielectron trigger has \pt
thresholds of 17 and 8\GeV on the leading and subleading electrons,
respectively. The dimuon trigger has thresholds that increased with
increased instantaneous luminosity, from 7\GeV on both muons to 13 and
8\GeV on the leading and subleading muons, respectively. Events from
pure beam-related backgrounds are rejected by requiring that at least
one primary vertex be reconstructed within the luminous region
($|z|<24$\,cm), with a fit based on at least five tracks.

Inclusive \ZGS+\,jets and \ttbar+\,jets events are simulated with
\MADGRAPH, using \PYTHIA for the parton shower, hadronisation, and
MPIs. The \ZGS boson is simulated with a minimum mass of 50\GeV and
only leptonic decays are considered. The \ZGS+\,jets sample is
normalised to the integrated luminosity in data using the cross
section of $3048 \pm 130$\,pb~\cite{Melnikov:2006}, which accounts for
the ${\mathcal{O}} (\alpha_S^2)$ next-to-next-to-leading-order (NNLO)
corrections to the inclusive \ZGS production. For the \ttbar+\,jets
sample the NLO cross section of $158^{+23}_{-24}$\,pb~\cite{TOPNLO} is
used. Pile-up events are added by assuming a flat distribution of
additional interaction vertices up to 10 vertices, and a tail from a
Poisson distribution above 10. All MC events are reweighted to
reproduce the number of pile-up events expected in data, as derived
from the instantaneous luminosity distribution.

Both electrons (muons) are required to be reconstructed with $\pt >
25$ (20)\GeV~\cite{EReco,MuReco}. The electrons (muons) are further
required to be well within the detector acceptance, with
pseudorapidity $|\eta| < 2.5$ (2.1). For electrons, the transition
region $1.444 < |\eta_{SC}| < 1.566$ between the barrel and endcap
parts of the ECAL is excluded, $\eta_{SC}$ being the pseudorapidity of
the electron ECAL cluster. Energy deposits from final state radiation
(FSR) are recombined with electrons during the reconstruction process
but not with muons. Nonprompt leptons are rejected by requiring a
maximum distance of closest approach between the track and the beam
axis in the transverse plane of 200\micron.

The lepton isolation is defined using the sum of transverse energy (or
\pt) around the lepton in a cone size $\DR < 0.3$ in the tracker
($I_{\mathrm{trk}}$), ECAL ($I_{\mathrm{ECAL}}$), and HCAL
($I_{\mathrm{HCAL}}$) detectors, with $\DR = \sqrt{(\Delta \eta)^2 +
  (\Delta \phi)^2}$. For electrons, separate selections are applied to
$I_{\mathrm{trk}}/\pt^{\mathrm{e}}$,
$I_{\mathrm{ECAL}}/\pt^{\mathrm{e}}$, and
$I_{\mathrm{HCAL}}/\pt^{\mathrm{e}}$. Electron identification and
isolation criteria are chosen to provide 85\% efficiency on the 2010
data sample~\cite{CMS_paper_WZ}, and are more stringent than the
requirements applied at the trigger level. Muon identification
criteria are mostly based on cosmic-ray rejection and the quality of
the global fit including the tracker and the muon chambers. For muon
isolation, the combined variable
$(I_{\mathrm{trk}}+I_{\mathrm{ECAL}}+I_{\mathrm{HCAL}})/\pt^{\mu}$ is
required to be less than 0.15. With this choice, the probability of
having two misidentified leptons is negligible. Opposite charges for
the leptons are required when forming pairs, and the lepton invariant
mass $M_{\ell\ell}$ is required to lie between 60 and 120\GeV. In the
case of multiple combinations, the lepton pair with the invariant mass
closest to the mass of the Z boson is kept.

Jets are reconstructed using the particle-flow (PF) objects that are
described in detail in
Refs.~\cite{CMS_PAS_PFT-09-001,CMS_PAS_PFT-10-001}.  Individual
particles (leptons, photons, charged and neutral hadrons) are
reconstructed by linking tracks, ECAL clusters, and HCAL clusters.
Each particle is reconstructed with the optimal momentum or energy
resolution by considering information from all subdetectors: charged
hadrons are reconstructed from tracks; photons and neutral hadrons are
reconstructed from energy clusters in the ECAL and HCAL. These
individual particles are then clustered into jets using the anti-k$_T$
jet clustering algorithm~\cite{Salam} with a distance parameter of
0.5, as implemented in the {\sc fastjet}
package~\cite{Cacciari:2005hq,FastJet}. Jets are calibrated to ensure
a uniform energy response in \pt and $\eta$, using photon+jet, Z+jet,
and dijet events~\cite{Chatrchyan:2011ds}. The contribution to the jet
energy from pile-up is estimated on an event-by-event basis using the
jet area method described in Ref.~\cite{cacciari-2008-659}, and is
subtracted from the overall $E_\mathrm{T}$ response.

The reconstructed jets are required to have $\pt > 25$\GeV and to be
separated from each of the \ZGS leptons by at least $\DR =0.5$.
Furthermore, jets are required to have $|\eta| < 2.1$, to ensure
optimal b-tagging performance. Loose identification
criteria~\cite{CMS_PAS_JME-10-001} are applied in order to further
reject jets coming from beam background, and to reject calorimeter
noise and isolated photons. These criteria are based on the
requirements that the total energy of the jet be shared between more
than one HCAL readout cell and not originate entirely from deposits
associated with neutral particles; the selection efficiency for
genuine jets is close to 100\% in both data and MC events.

The efficiency of the dilepton selection is estimated in data and MC
simulation using the tag-and-probe method introduced in
Ref.~\cite{CMS_paper_WZ} in events with at least two leptons and a jet
passing the requirements detailed above. For data-MC comparisons, MC
events are reweighted according to the data/MC scale factors per
lepton, as a function of their \pt and $\eta$.  The systematic
uncertainty on the scale factor per lepton is less than 2\% for
electrons, and less than 1\% for muons.

The \ZGS+\,jets MC sample is split into three subsamples, according to
the underlying production of b jets, c jets, or jets originating only
from gluons or u, d, s quarks (hereafter called light jets), with no
requirement on the \pt or $\eta$ of the jets.  These subsamples are
labelled respectively Z+b, Z+c, and Z+l. The fraction of events after
the dilepton+jet selection that contain at least one reconstructed jet
with $\pt > 25$\GeV and $|\eta| < 2.1$, matched to a generator-level b
(c) quark in \DR, is about 4\% (7\%).

\begin{figure*}[h!]
  \begin{center}
    \includegraphics[width=0.48\textwidth]{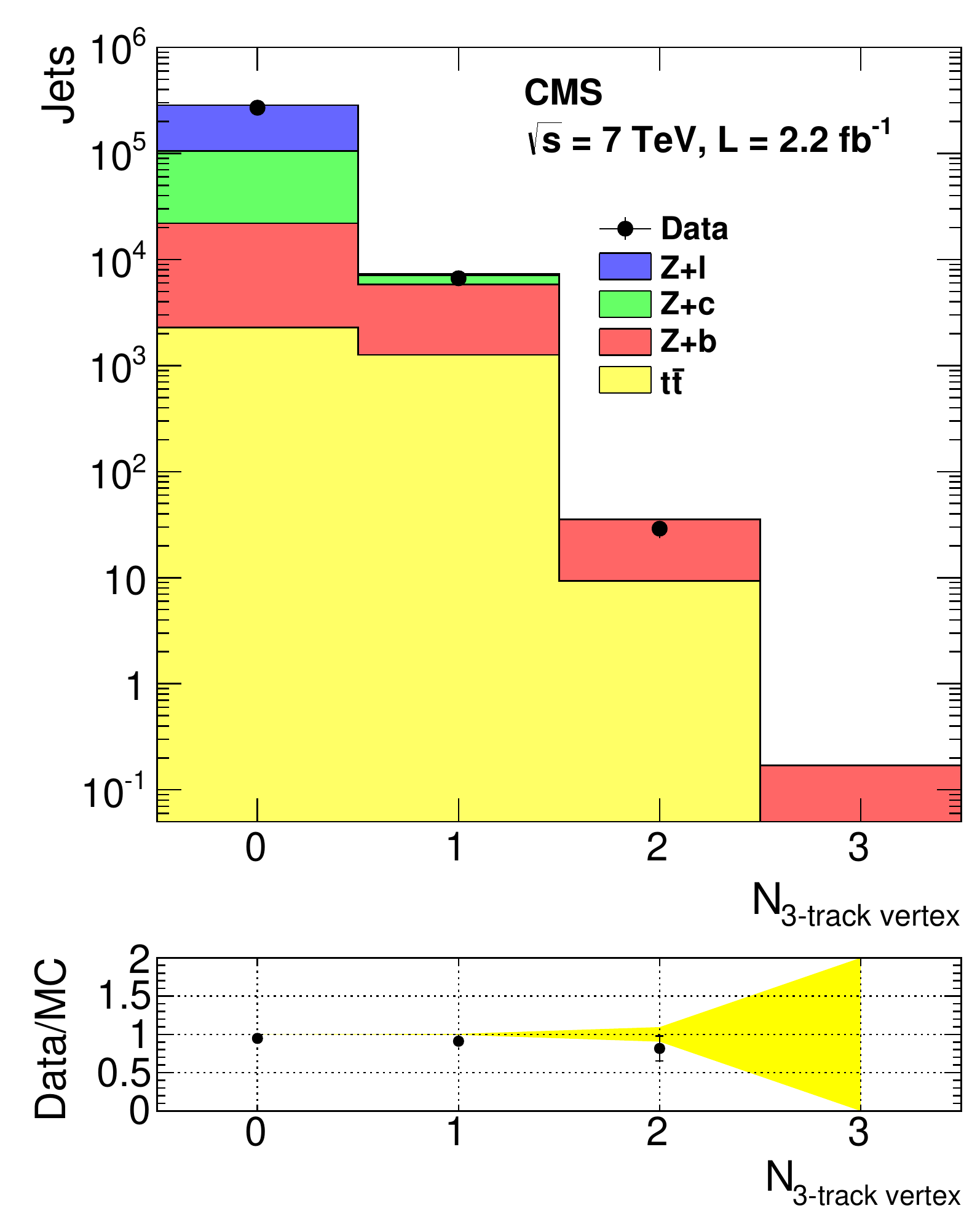}
    \includegraphics[width=0.48\textwidth]{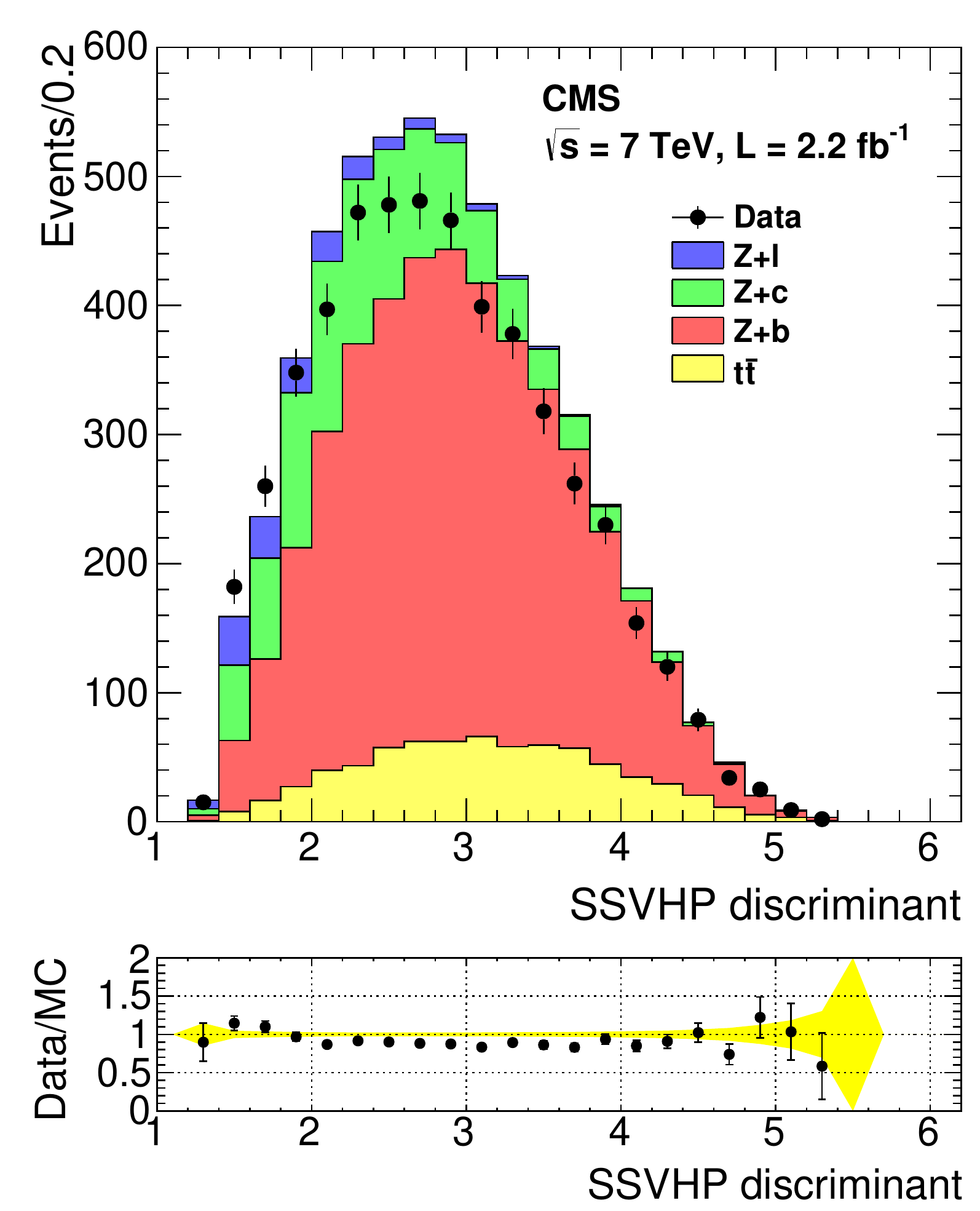}
    \caption{Left: number of three-track secondary vertices per jet in
      dilepton+jet events. Right: SSV HP discriminant for the leading
      jet after the dilepton+jet selection. The yellow bands in the
      lower plots represent the statistical uncertainty on the MC
      yield.}
    \label{fig:ssv_jetSel}
  \end{center}
\end{figure*}

Jets originating from b quarks are tagged by taking advantage of the
long b-hadron lifetime.  The Simple Secondary Vertex (SSV) algorithm
discriminant is a monotonic function of the three-dimensional flight
distance significance from the reference primary vertex (i.e.\ the
primary vertex with the highest quadratic sum of the \pt of its
constituent tracks, $\sum_{\mathrm{tracks}}\pt^2$) to the chosen
secondary vertex. Values of the discriminant greater than one indicate
the presence of a secondary vertex. To improve the purity of the
selection, only secondary vertices built from at least three tracks
are considered, referred to as high purity (HP) vertices in the
following. The number of HP secondary vertices per jet is shown in
Fig.~\ref{fig:ssv_jetSel} (left), after the dilepton+jet
selection. The leading jet is found to have at least one HP secondary
vertex in $2.4 \pm 0.2$\% of the dilepton+jet events. The distribution
of the SSV HP discriminant is shown for the leading jet in
Fig.~\ref{fig:ssv_jetSel} (right). The discriminant value to define b
jets is chosen to be 2.0, such that the rate of tagging a light jet
(mistagging rate) is below 0.1\%. Further details can be found in
Refs.~\cite{CMS_PAS_BTV-10-001,CMS_PAS_BTV-11-002}.

The b-tagging efficiencies and mistagging rates are measured in the
data and MC samples, as functions of the \pt and $\eta$ of the jet,
using inclusive jet samples, as described in
Ref.~\cite{CMS_PAS_BTV-11-001}. The tagging efficiency in the MC
simulation is found to be higher than that in data, as can be seen in
Fig.~\ref{fig:ssv_jetSel} (right) in the discrepancy between data and
MC observed in the discriminant above 2.0.  In all subsequent results,
a weight is applied to the MC events to reproduce the b-tagging
efficiency and mistagging rate measured in the data. This weight takes
into account the appropriate data/MC scale factor for each b-tagged
jet, depending on the generator-level flavour. The MC b-jet efficiency
is extracted from the signal Z+b MC simulation.

\begin{figure*}[h!]
  \begin{center}
    \includegraphics[width=0.48\textwidth]{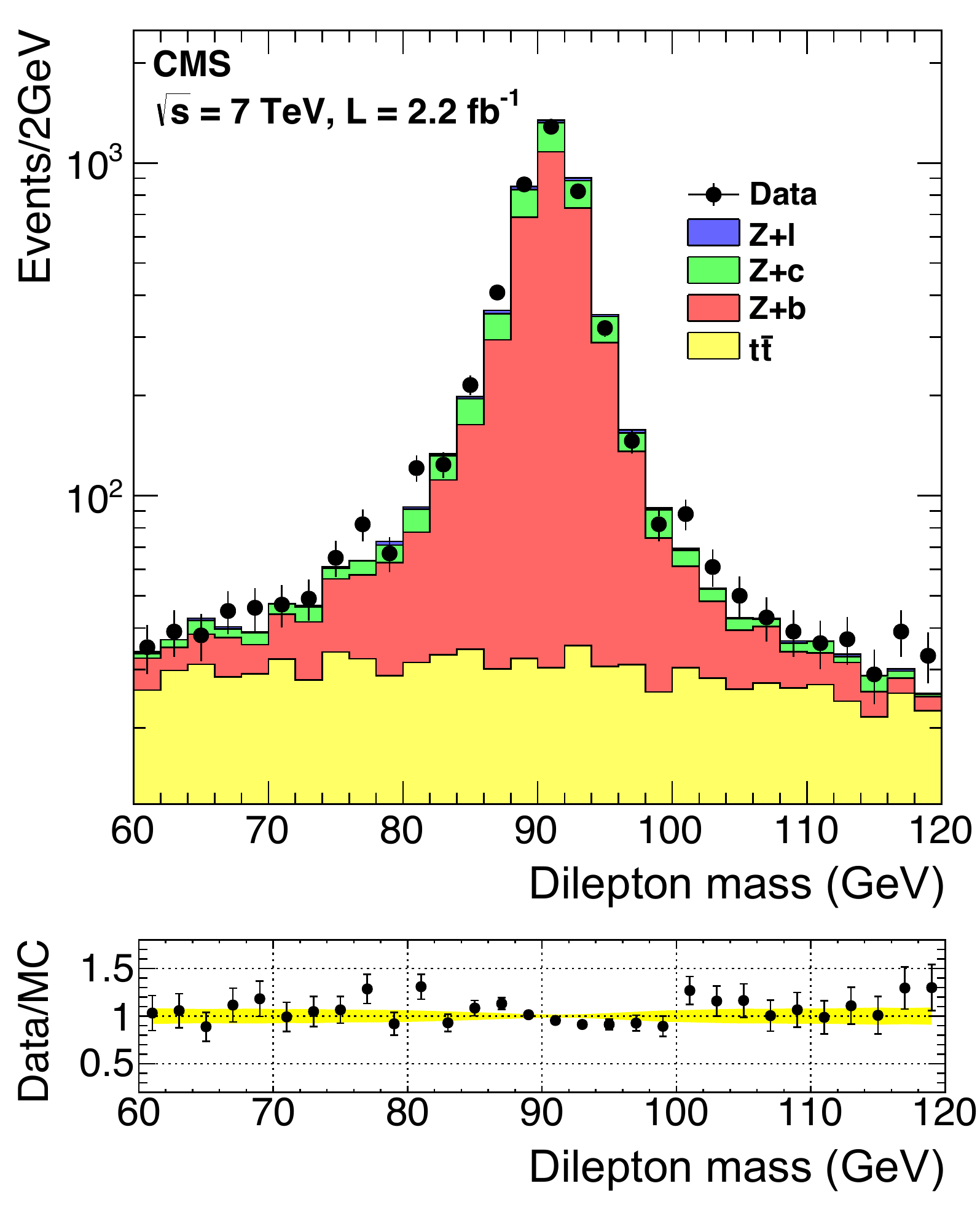}
    \includegraphics[width=0.48\linewidth]{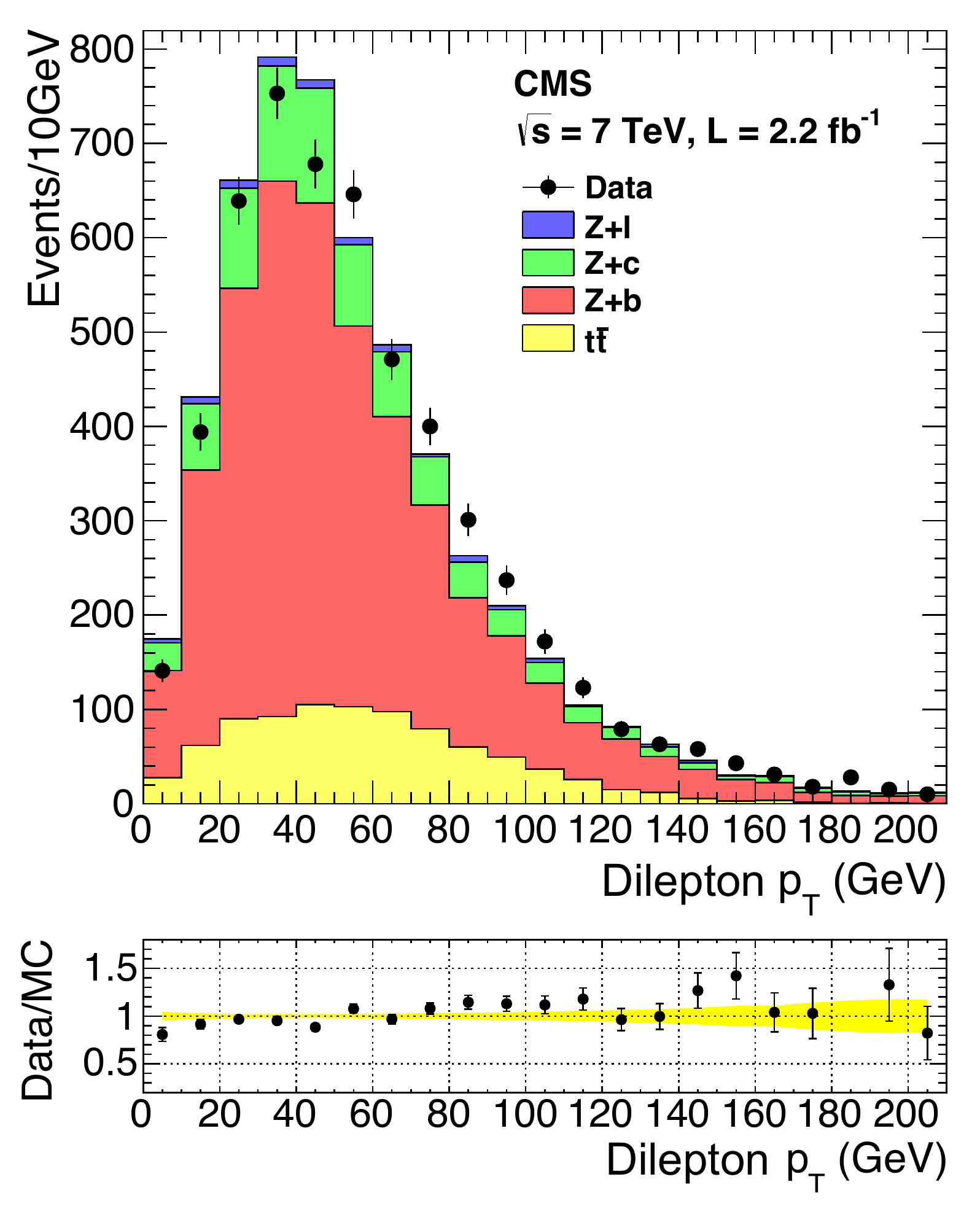}
    \caption{Invariant mass (left) and \pt (right) of the lepton pairs
      after the dilepton+b-jet selection. The yellow bands in the
      lower plots represent the statistical uncertainty on the MC
      yield.  }
    \label{fig:kinll_bjetSel}
  \end{center}
\end{figure*}

Before (after) b tagging, the reference vertex is found to be
identical to the dilepton vertex in more than 99.7\% (99.9\%) of the
events. Therefore, no explicit requirement of a common vertex is
applied to the dilepton and b jet.  The invariant mass and \pt
distributions of the lepton pairs are shown in
Fig.~\ref{fig:kinll_bjetSel} after the dilepton+b-jet selection. A
discrepancy between data and MC simulation is observed in the dilepton
$\pt$ distribution, especially in the region between 50 and 120\GeV.
This hardening of the spectrum in data could come from higher-order
corrections~\cite{Alwall:2007fs,Alwall:2008qv}.

\begin{figure*}[h!]
  \begin{minipage}{0.48\linewidth}
    \includegraphics[width=1.\linewidth]{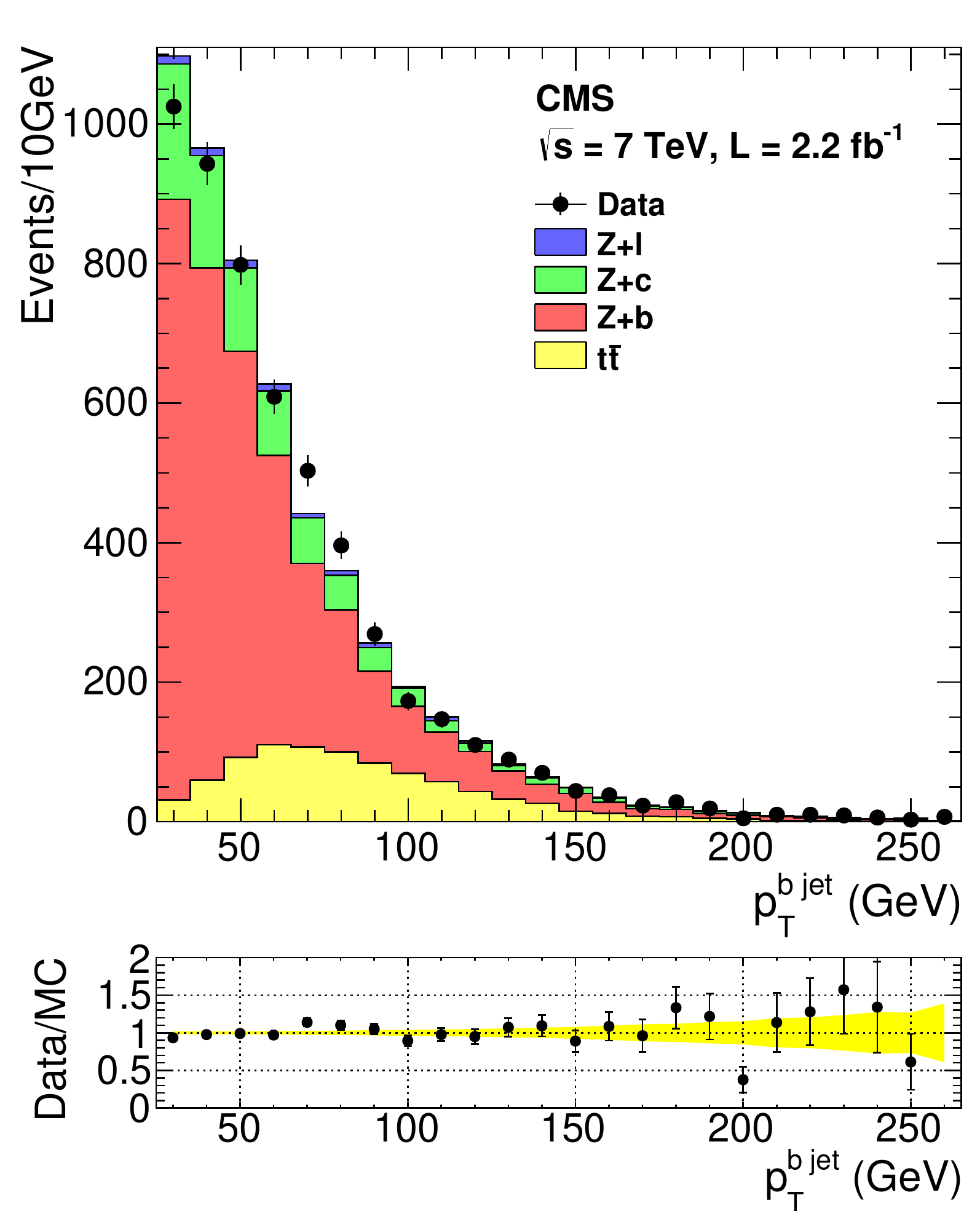}
  \end{minipage}
  \hfill
  \begin{minipage}{0.48\linewidth}
    \includegraphics[width=1.\linewidth]{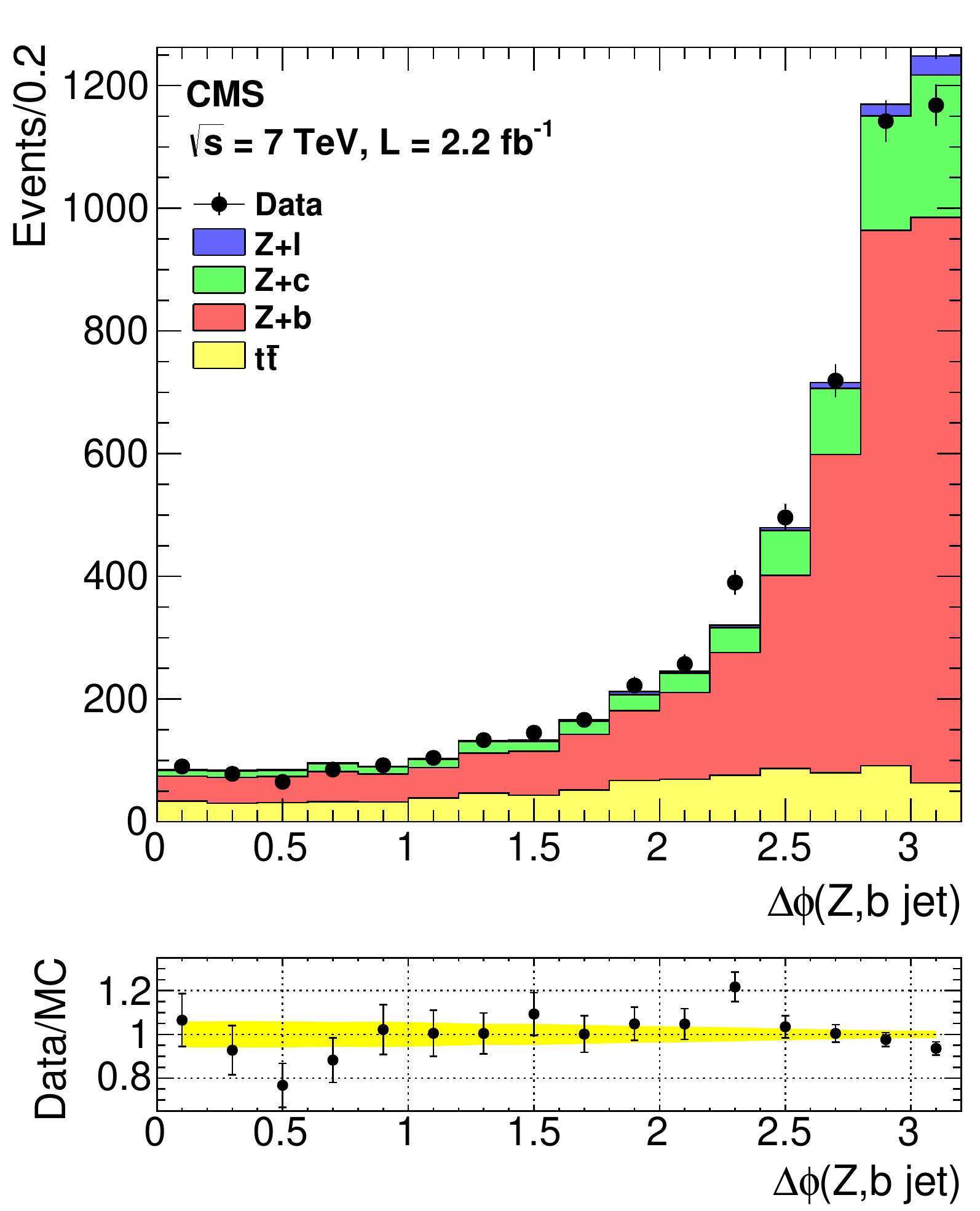}
   \end{minipage}
   \caption{Left: \pt of the leading-\pt b jet after the
     dilepton+b-jet selection. Right: distribution of $\Delta
     \phi$(Z,\,b jet) between the leading-\pt b jet and the lepton
     pair after the dilepton+b-jet selection. The yellow bands in the
     lower plots represent the statistical uncertainty on the MC
     yield.  }
    \label{fig:kinb_bHPSel}
\end{figure*}

The \pt distribution for the b-tagged jet with the largest transverse
momentum, hereafter called the leading-\pt b jet, is shown after the
dilepton+b-jet selection in Fig.~\ref{fig:kinb_bHPSel} (left).  A
deficit in MC events is seen at around 70\GeV. The distribution of
$\Delta \phi$(Z,\,b jet) between the leading-\pt b jet and the lepton
pair is shown in Fig.~\ref{fig:kinb_bHPSel} (right). A deficit in MC
events is seen in the region $2 < \Delta \phi(\mbox{Z,\,b jet})<2.7$.

The cross section for the production of a \ZGS boson in association
with at least one hadron-level b jet is extracted from the selected
numbers of dielectron+b-jet and dimuon+b-jet events, taking into
account the b-jet purity $\mathcal{P}$, the fraction $f_{\ttbar}$ of
$\ttbar$ events, the b-tagging efficiency $\varepsilon_{\mathrm{b}}$,
the lepton efficiency $\varepsilon_{\ell}$, the correction factor
$\mathcal{C}_{\mathrm{hadron}}$ for detector and reconstruction
effects, and the lepton acceptance $\mathcal{A}_{\ell}$, using the
following equation:

\begin{equation}
\label{eq:xs_formula}
\sigma_{\mathrm{hadron}}(\ZGS+\mathrm{b},\ZGS\rightarrow \ell\ell) = \frac{N(\ell\ell+\mathrm{b})\times (\mathcal{P} - f_{\ttbar})}{\mathcal{A}_{\ell} \times \mathcal{C}_{\mathrm{hadron}} \times  \varepsilon_{\ell} \times \varepsilon_{\mathrm{b}} \times \mathcal{L}}\ .
\end{equation}

It is defined with the following requirements: (i) $\pt^{\mathrm{b}} >
25$\GeV and $|\eta^{\mathrm{b}}| < 2.1$ on any hadron jet containing a
b hadron, (ii) $60 < M_{\ell\ell} < 120$\GeV, and (iii)
$\DR(\mbox{jet,\,leptons})>0.5$.

\begin{figure*}[h!]
  \includegraphics[width=0.48\textwidth]{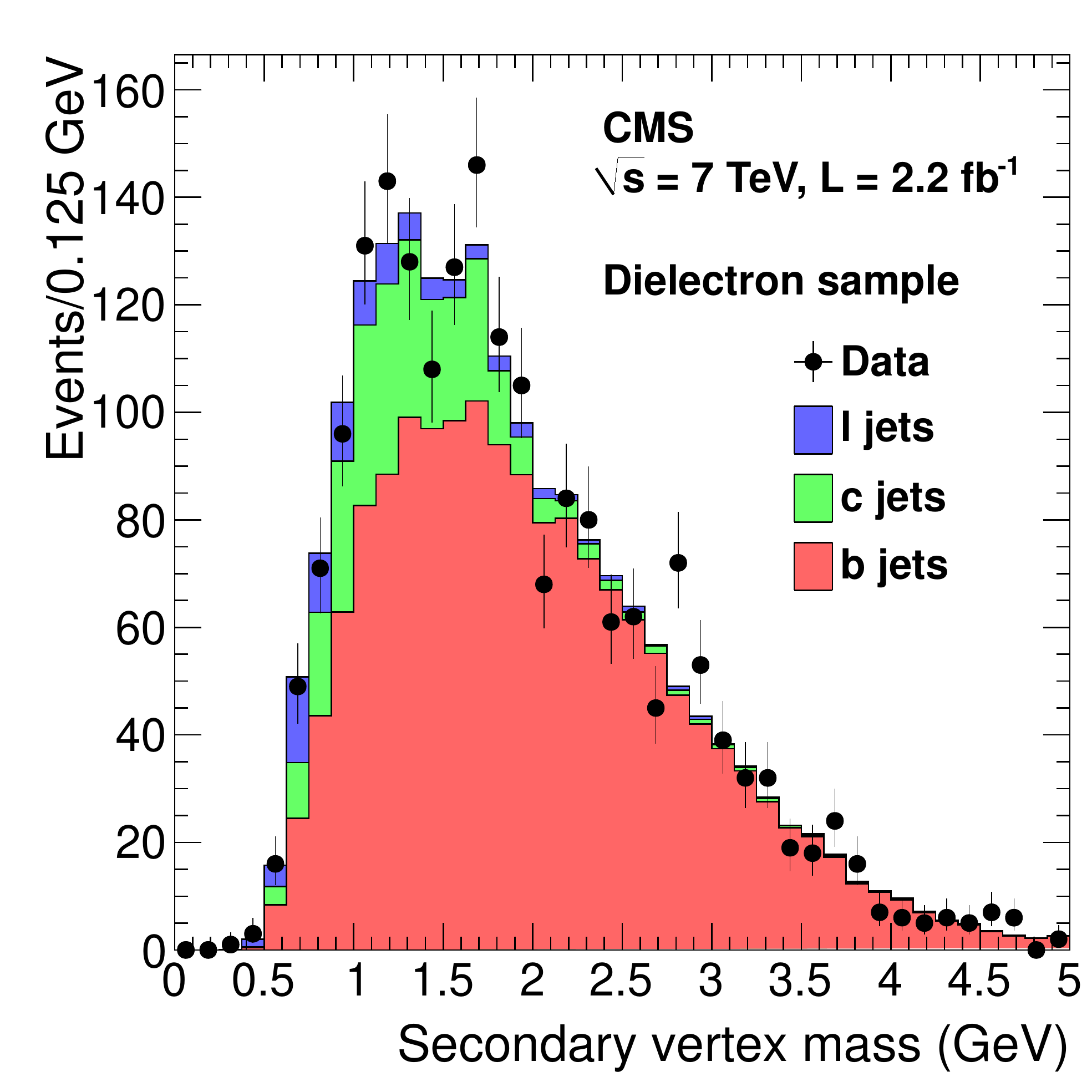}
  \hfill
  \includegraphics[width=0.48\textwidth]{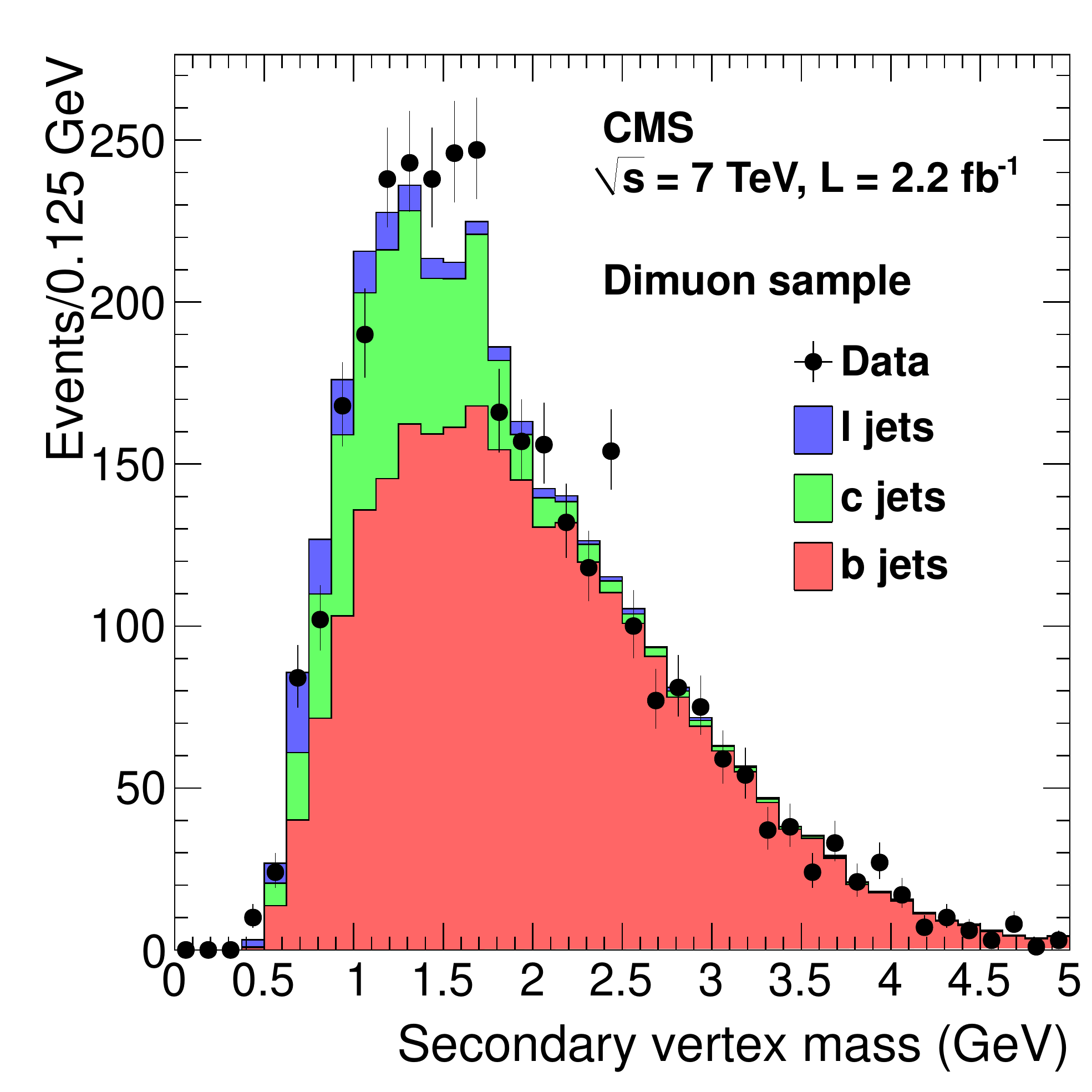}
  \caption{Secondary-vertex mass of the leading-\pt b jet, with MC
    distributions constructed from inclusive jet samples, for the
    dielectron (left) and dimuon (right) channels, after the b-purity
    maximum-likelihood fit.}
    \label{fig:ssvmass_bjetSel}
\end{figure*}

The event-level b-jet purity $\mathcal{P}$ is extracted by means of a
fit to the distribution of secondary-vertex mass of the leading-\pt b
jet in the data. The mass of the secondary vertex is defined as the
invariant mass of all tracks originating from the secondary vertex,
assuming the pion mass for each track. Separate sets of distributions
for b, c, and light jets are derived from the MC simulation: (i) using
the \ZGS+\,jets sample and (ii) using inclusive jet samples reweighted
to match the \pt and $\eta$ spectra of the leading-\pt candidate jet
in the dilepton+b-jet datasets.  While the distributions from the
inclusive jet samples were used as a baseline, the systematic
uncertainty on the purity is calculated from the differences between
the two sets and from the statistical uncertainty from the fit. The
fraction of events in which the b hadron originates from gluon
splitting is found to have a negligible impact on the shapes, but is
nevertheless included in the shape uncertainty. The secondary-vertex
mass is shown in Fig.~\ref{fig:ssvmass_bjetSel} for the dielectron
(left) and dimuon (right) channels.

As \ttbar production also yields genuine b jets, \ttbar events must be
subtracted after the purity correction. The fraction $f_{\ttbar}$ of
\ttbar background events remaining after the selection is evaluated
from the data using the dilepton invariant mass distribution. The
\ttbar contribution in the region of the Z-boson mass peak, $[60\GeV,
120\GeV]$, is extrapolated from the upper sideband region, which is
dominated by \ttbar events. The ratio of the numbers of \ttbar events
in the two regions is taken from the MC simulation and corrected for
discrepancies between data and simulation using the dileptonic \ttbar
decay channel in the background-free e$\mu$bb final
state. Consequently, the systematic uncertainty on the \ttbar
contribution is dominated by the uncertainty on the data/MC scale
factor obtained with the dileptonic \ttbar decay
measurement.

Other backgrounds, e.g.~diboson and single-top processes, yield
negligible contributions which are mostly removed after taking into
account the purity. The MPI effects are included in the measurement
and are expected to contribute to the cross section at a level of
about 2\%~\cite{dzeroMPI}.

The average efficiency $\varepsilon_{\ell}$ of the dilepton selection
is evaluated using the MC Z+b signal sample corrected event-by-event
to reproduce the lepton efficiencies measured in data, as explained
previously. Systematic uncertainties on $\varepsilon_{\ell}$ arise
from the tag-and-probe analysis.

Similarly, the average efficiency $\varepsilon_{\mathrm{b}}$ of the
b-jet selection is evaluated using the MC Z+b signal sample, corrected
event-by-event to reproduce the b-tagging efficiency and mistagging
rates measured in the data, as explained previously. Systematic
uncertainties arise mainly from the uncertainty on the weights
applied: (i) the data/MC scale factor uncertainty of 10\% for b and c
jets, and 10--20\% for light jets; (ii) the mistagging rate
uncertainty of 20--30\%. These uncertainties are documented in
Refs.~\cite{CMS_PAS_BTV-11-001,CMS_PAS_BTV-11-002}.

The correction factor $\mathcal{C}_{\mathrm{hadron}}$ is introduced to
account for detector resolution and other reconstruction effects. It
is computed from the MC \MADGRAPH+\PYTHIA signal sample by comparing
the event yields at detector level to the event yields at generator
level. The same kinematic selections as for the reconstructed leptons
and jets are applied to the generator-level objects (including the
selection in $M_{\ell\ell}$ and \DR(jet,\,leptons)). To ensure that
the correct objects are selected at the detector level, only jets
matched to generator-level b jets and leptons matched to the
generator-level \ZGS-decay leptons are considered. Hadron jets are
defined using the anti-k$_T$ clustering algorithm with a distance
parameter of 0.5 applied to all stable particles but neutrinos after
the hadronisation. A hadron jet is labelled as a b jet if there is a b
hadron within $\DR=0.5$ of the jet axis. Systematic uncertainties on
$\mathcal{C}_{\mathrm{hadron}}$ are derived by using \SHERPA and
{\footnotesize{a}}\MCATNLO+\HERWIG, and from the uncertainty on the
jet energy resolution (14\% for jets in the barrel, 22\% in the
endcaps)~\cite{Chatrchyan:2011ds}. In the electron case,
$\mathcal{C}_{\mathrm{hadron}}$ contains a small acceptance term
coming from the ECAL transition region being removed at the
reconstructed level. In the muon case, it contains a correction from
FSR.

Once the cross section is corrected back to the particle level, a
final acceptance factor $\mathcal{A}_{\ell}$ is applied to correct for
the efficiency of the lepton acceptance selection:
\ZGS$\rightarrow\Pe\Pe$ with $\pt^{\mathrm{e}} > 25$\GeV and
$|\eta^{\mathrm{e}}| < 2.5$, or \ZGS$\rightarrow \mu\mu$ with
$\pt^{\mu} > 20$\GeV and $|\eta^{\mu}| < 2.1$, the electron and muon
properties being defined before FSR. The systematic uncertainty on
$\mathcal{A}_{\ell}$ is evaluated with \SHERPA,
{\footnotesize{a}}\MCATNLO+\HERWIG, and
MCFM~\cite{Campbell:2003dd,Maltoni:2005wd,Campbell:2005zv}.

The ($\pt$,$\eta$)-dependent jet-energy-scale uncertainty amounts to
3--5\% of the \pt of the jet. Its effect on the cross section is
estimated to be 2.5\% using the MC signal sample, reweighted to match
the data. To estimate the uncertainty due to the pile-up, the mean of
the expected distribution used to reweight the MC simulation is
shifted up and down by 0.6 interactions.

The estimates of the parameters defined above and the resulting cross
sections are summarised in Table~\ref{tab:xsTerms} for the $\Pe\Pe$+b
and $\mu\mu$+b selections. The contributions expected from the Z+b MC
signal sample are $1308 \pm 15 \stat$ and $2078 \pm 19 \stat$ events
for the ee+b and $\mu\mu$+b selections, respectively, to be compared
with the background-corrected data yields of $1288 \pm 29 \stat \pm 84
\syst$ and $2121 \pm 37 \stat \pm 124 \syst$ events. The theoretical
uncertainties on the cross section results presented in
Table~\ref{tab:xsTerms} come from the systematic uncertainties on
$\mathcal{C}_{\mathrm{hadron}}$ and $\mathcal{A}_{\ell}$ that were
estimated using different MC models.  Fractional uncertainties on the
cross section results are summarised in Table~\ref{tab:systematics}.

\begin{table*}[h!]
\newcommand\T{\rule{0pt}{2.6ex}}
\newcommand\B{\rule[-1.2ex]{0pt}{0pt}}

\begin{center}
  \caption{Extraction of the cross section
    $\sigma_{\mathrm{hadron}}(\ZGS+\mathrm{b},\ZGS\rightarrow
    \ell\ell)$ for $\ell\ell =\Pe\Pe$ or $\mu\mu$. The uncertainty on
    each parameter contains all the systematic effects considered in
    the analysis, summarised in Table~\ref{tab:systematics} and
    detailed in the main text. The first uncertainty on the cross
    section results is statistical, while the second is systematic,
    and the third accounts for limitations of the
    theory.\label{tab:xsTerms}}
\begin{tabular}{|l|c|c|}
  \hline
  Variable     & $\Pe\Pe$+b & $\mu\mu$+b \\
  \hline
  Selected events & $1990$ & $3362$ \\
  $\mathcal{P}$ (\%) & $83.4 \pm 3.6$ & $81.5 \pm 2.9$ \\
  $f_{\ttbar}$ (\%)   & $18.7 \pm 2.2$ & $18.4 \pm 2.3$ \\
$\varepsilon_{\mathrm{b}}$ (\%) & $35.3 \pm 3.5$ & $34.9 \pm 3.5$ \\
$\varepsilon_{\ell}$ (\%) & $63.2 \pm 2.6$ & $84.4 \pm 1.7$ \\
\hline
$\mathcal{C}_{\mathrm{hadron}}$ (\%)\T \B  & $84.2^{+5.8}_{-0.6} $ & $95.0 ^{+6.6}_{-0.5} $ \\
$\mathcal{A}_{\ell}$ (\%)\T \B  & $55.0 _{-2.1}^{+3.6}$ & $57.2 _{-2.4}^{+3.7}$ \\
\hline
$\sigma_{\mathrm{hadron}}(\ZGS+\mathrm{b},\ZGS\rightarrow \ell\ell)$ (pb)\T \B  & $5.61 \pm 0.13 \pm 0.73 ^{+0.24}_{-0.53}$ & $5.97 \pm 0.10 \pm 0.73 ^{+0.25}_{-0.57}$ \\
\hline
\end{tabular}
\end{center}
\end{table*}

 \newcolumntype{x}[1]{%
>{\centering\arraybackslash\hspace{0pt}}p{#1}}%

\begin{table*}[h!]
\newcommand\T{\rule{0pt}{2.6ex}}
\newcommand\B{\rule[-1.2ex]{0pt}{0pt}}
\begin{center}
  \caption{Fractional uncertainties on the cross section measurement
    from the different sources considered.\label{tab:systematics}}
\begin{tabular}{|l|x{0.15\textwidth}|x{0.15\textwidth}|}
\hline
Correlated sources & \multicolumn{2}{c|}{Fractional uncertainty (\%)} \\
\hline
b-tagging efficiency  & \multicolumn{2}{c|}{10} \\
b-jet purity & \multicolumn{1}{c}{5.6 ($\Pe\Pe$+b)} & 4.6 ($\mu\mu$+b) \\
\ttbar contribution & \multicolumn{2}{c|}{2.9} \\
Jet energy scale & \multicolumn{2}{c|}{2.5} \\
Luminosity & \multicolumn{2}{c|}{2.2} \\
Jet energy resolution & \multicolumn{2}{c|}{0.5}  \\
Pile-up & \multicolumn{1}{c}{1.5 ($\Pe\Pe$+b)} & 0.5 ($\mu\mu$+b) \\
Mistagging rate & \multicolumn{2}{c|}{0.04} \\
\hline
Theory (via ${\mathcal A}_{\ell}$)\T \B  & \multicolumn{2}{c|}{$_{-6.5}^{+4.2}$} \\
Theory (via $\mathcal{C}_{\mathrm{hadron}}$)\T \B  & \multicolumn{2}{c|}{$_{-6.9}^{+0.7}$} \\
\hline
\hline
Uncorrelated sources  & $\Pe\Pe$+b & $\mu\mu$+b \\
\hline
Trigger and dilepton selection  & 4 & 2 \\
\ttbar contribution & 1.9 & 2.2 \\
\hline
\hline
Experimental systematic & 13.0 & 12.3 \\
Theoretical systematic\T \B   & $_{-9.5}^{+4.2}$ &  $_{-9.5}^{+4.2}$ \\
Statistical  & 2.2  & 1.7 \\
\hline
\end{tabular}
\end{center}
\end{table*}

After correction for the b-tagging efficiencies and the lepton
acceptance requirements, results for the $\Pe\Pe$ and $\mu\mu$ selections
are found to be in good agreement, $5.61 \pm 0.13 \stat \pm 0.73 \syst$
$^{+0.24}_{-0.53} \theory$\,pb and $5.97 \pm 0.10\stat \pm 0.73 \syst
^{+0.25}_{-0.57} \theory$\,pb, respectively. The $\Pe\Pe$ and $\mu\mu$
results are combined, taking into account correlated uncertainties as
given in Table~\ref{tab:systematics}, and the final result is found to
be $5.84 \pm 0.08 \stat \pm 0.72 \syst _{-0.55}^{+0.25} \theory$\,pb.

The results are compared to the NLO calculations obtained with the
MCFM tool~\cite{Campbell:2003dd,Maltoni:2005wd,Campbell:2005zv}. The
inclusive cross section at parton level is found to be
$\sigma_{\mathrm{parton}}^{\mathrm{MCFM}} = 4.73 \pm 0.54$\,pb, using
the same acceptance requirements for the leptons and parton jets. The
uncertainty on the MCFM estimate comes from the CTEQ6M PDF
set~\cite{Pumplin:2002vw} and variations of the renormalisation and
factorisation scales by factors of 0.5 and 2 around the mass of the Z
boson, considering both correlated and anticorrelated
combinations~\cite{PDF4LHC}. In order to extract the corresponding
prediction at the hadron level, nonperturbative (NP) effects like
hadronisation are quantified. A correction factor $C_{\mathrm{NP}}$ is
computed from parton to hadron level using \MADGRAPH+\PYTHIA and
{\footnotesize{a}}\MCATNLO+\HERWIG. Parton jets are defined using the
anti-k$_T$ clustering algorithm with a distance parameter of 0.5,
applied to all quarks and gluons after showering but before
hadronisation. A parton jet is labelled as a b jet if there is a b
quark among its constituents. The correction is found to be
$C_{\mathrm{NP}} = (84 \pm 3)\%$, leading to a hadron-level-corrected
NLO prediction of $3.97 \pm 0.47$\,pb. The theoretical prediction in
the context of this MCFM calculation is found to be smaller than the
data measurement.

In conclusion, the production of b jets in association with a \ZGS
boson has been studied in 2.2\,fb$^{-1}$ of proton-proton collision
data at a centre-of-mass energy of 7\TeV recorded by the CMS
detector. Measurements performed in the electron and muon \ZGS-decay
channels are combined. The \MADGRAPH simulation interfaced with
\PYTHIA is used to derive the correction from the reconstructed level
to the hadron level. The \ZGSB cross section, with
$\ZGS\rightarrow\ell\ell$ where $\ell\ell=\Pe\Pe$ or $\mu\mu$, for
events with the \ZGS lepton pair invariant mass $60 < M_{\ell\ell} <
120$\GeV, and at least one b jet at the hadron level with \pt $>
25$\GeV and $|\eta|<2.1$, using anti-k$_T$ jets reconstructed with a
distance parameter of 0.5 and with a separation between the leptons
and the jets of $\DR > 0.5$, is found to be $5.84 \pm 0.08 \stat \pm
0.72 \syst _{-0.55}^{+0.25} \theory$\,pb. The distributions of the
kinematic variables for the leading-\pt b jet and the \ZGS-decay
leptons are found to be in fair agreement with the predictions made by
the \MADGRAPH event generator interfaced with \PYTHIA, and normalised
to the integrated luminosity in data using a cross section value that
includes the NNLO corrections to the inclusive \ZGS production. The
residual discrepancy may be a consequence of the higher order terms
absent in the \MADGRAPH tree-level simulation in the variable-flavour
scheme with massless b quarks.

We congratulate our colleagues in the CERN accelerator departments for
the excellent performance of the LHC machine. We thank the technical
and administrative staff at CERN and other CMS institutes, and
acknowledge support from: FMSR (Austria); FNRS and FWO (Belgium);
CNPq, CAPES, FAPERJ, and FAPESP (Brazil); MES (Bulgaria); CERN; CAS,
MoST, and NSFC (China); COLCIENCIAS (Colombia); MSES (Croatia); RPF
(Cyprus); MoER, SF0690030s09 and ERDF (Estonia); Academy of Finland,
MEC, and HIP (Finland); CEA and CNRS/IN2P3 (France); BMBF, DFG, and
HGF (Germany); GSRT (Greece); OTKA and NKTH (Hungary); DAE and DST
(India); IPM (Iran); SFI (Ireland); INFN (Italy); NRF and WCU (Korea);
LAS (Lithuania); CINVESTAV, CONACYT, SEP, and UASLP-FAI (Mexico); MSI
(New Zealand); PAEC (Pakistan); MSHE and NSC (Poland); FCT (Portugal);
JINR (Armenia, Belarus, Georgia, Ukraine, Uzbekistan); MON, RosAtom,
RAS and RFBR (Russia); MSTD (Serbia); MICINN and CPAN (Spain); Swiss
Funding Agencies (Switzerland); NSC (Taipei); TUBITAK and TAEK
(Turkey); STFC (United Kingdom); DOE and NSF (USA). Individuals have
received support from the Marie-Curie programme and the European
Research Council (European Union); the Leventis Foundation; the
A. P. Sloan Foundation; the Alexander von Humboldt Foundation; the
Belgian Federal Science Policy Office; the Fonds pour la Formation \`a
la Recherche dans l'Industrie et dans l'Agriculture (FRIA-Belgium);
the Agentschap voor Innovatie door Wetenschap en Technologie
(IWT-Belgium); the Council of Science and Industrial Research, India;
and the HOMING PLUS programme of Foundation for Polish Science,
cofinanced from European Union, Regional Development Fund.

\bibliography{auto_generated}   

\providecommand{\href}[2]{#2}\begingroup\raggedright\begin{thebibliography}{10}%
\makeatletter
\providecommand{\hrefCMSnoop }[0]{\@secondoftwo}%
\makeatother
\providecommand{\doi}{\texttt{doi:}\begingroup \urlstyle{tt}\Url}

\bibitem{Frederix:2011qg}
R.~Frederix\hrefCMSnoop {} { {et~al.}, ``{W and $Z/\gamma*$ boson production in
  association with a bottom-antibottom pair}'',} \textit{ JHEP} \textbf{ 09}
  (2011) 061,
  \href{http://dx.doi.org/10.1007/JHEP09(2011)061}{\doi{10.1007/JHEP09(2011)061}},
  \href{http://www.arXiv.org/abs/1106.6019}{\texttt{ arXiv:1106.6019}}.

\bibitem{Campbell:2003dd}
J.~M. Campbell\hrefCMSnoop {} { {et~al.}, ``{Associated production of a $Z$
  Boson and a single heavy quark jet}'',} \textit{ Phys. Rev. D} \textbf{ 69}
  (2004) 074021,
  \href{http://dx.doi.org/10.1103/PhysRevD.69.074021}{\doi{10.1103/PhysRevD.69.074021}},
  \href{http://www.arXiv.org/abs/hep-ph/0312024v2}{\texttt{
  arXiv:hep-ph/0312024v2}}.

\bibitem{Maltoni:2005wd}
\hrefCMSnoop {} {F.~Maltoni, T.~McElmurry, and S.~Willenbrock, ``{Inclusive
  production of a Higgs or $Z$ boson in association with heavy quarks}'',}
  \textit{ Phys. Rev. D} \textbf{ 72} (2005) 074024,
  \href{http://dx.doi.org/10.1103/PhysRevD.72.074024}{\doi{10.1103/PhysRevD.72.074024}},
  \href{http://www.arXiv.org/abs/hep-ph/0505014v3}{\texttt{
  arXiv:hep-ph/0505014v3}}.

\bibitem{Campbell:2005zv}
J.~M. Campbell\hrefCMSnoop {} { {et~al.}, ``{Production of a $Z$ boson and two
  jets with one heavy-quark tag}'',} \textit{ Phys. Rev. D} \textbf{ 73} (2006)
  054007,
  \href{http://dx.doi.org/10.1103/PhysRevD.73.054007}{\doi{10.1103/PhysRevD.73.054007}},
  \href{http://www.arXiv.org/abs/hep-ph/0510362v2}{\texttt{
  arXiv:hep-ph/0510362v2}}.

\bibitem{MADGRAPH5}
J.~Alwall\hrefCMSnoop {} { {et~al.}, ``{MadGraph 5: going beyond}'',} \textit{
  JHEP} \textbf{ 06} (2011) 128,
  \href{http://dx.doi.org/10.1007/JHEP06(2011)128}{\doi{10.1007/JHEP06(2011)128}},
  \href{http://www.arXiv.org/abs/1106.0522v1}{\texttt{ arXiv:1106.0522v1}}.

\bibitem{PYTHIA}
\hrefCMSnoop {} {{T. Sj\"ostrand, S. Mrenna, and P. Z. Skands}, ``{PYTHIA 6.4
  Physics and Manual}'',} \textit{ JHEP} \textbf{ 05} (2006) 026,
  \href{http://dx.doi.org/10.1088/1126-6708/2006/05/026}{\doi{10.1088/1126-6708/2006/05/026}},
  \href{http://www.arXiv.org/abs/hep-ph/0603175}{\texttt{
  arXiv:hep-ph/0603175}}.

\bibitem{Field:2010bc}
\hrefCMSnoop {} {R.~Field, ``Early {LHC} Underlying Event Data -- Findings and
  Surprises'',} (2010). \href{http://www.arXiv.org/abs/1010.3558}{\texttt{
  arXiv:1010.3558}}.

\bibitem{QCD-10-010}
\hrefCMSnoop {} {{ CMS} Collaboration, ``{Measurement of the Underlying Event
  Activity at the LHC with $\sqrt{s}= 7$\,TeV and Comparison with $\sqrt{s} =
  0.9$\,TeV}'',} \textit{ JHEP} \textbf{ 09} (2011) 109,
  \href{http://dx.doi.org/10.1007/JHEP09(2011)109}{\doi{10.1007/JHEP09(2011)109}},
  \href{http://www.arXiv.org/abs/1107.0330v1}{\texttt{ arXiv:1107.0330v1}}.

\bibitem{Gleisberg:2008ta}
T.~Gleisberg\hrefCMSnoop {} { {et~al.}, ``{Event generation with SHERPA
  1.1}'',} \textit{ JHEP} \textbf{ 02} (2009) 007,
  \href{http://dx.doi.org/10.1088/1126-6708/2009/02/007}{\doi{10.1088/1126-6708/2009/02/007}},
  \href{http://www.arXiv.org/abs/0811.4622}{\texttt{ arXiv:0811.4622}}.

\bibitem{Marchesini:1991ch}
G.~G.~Marchesinia\hrefCMSnoop {} { {et~al.}, ``{HERWIG: A Monte Carlo event
  generator for simulating hadron emission reactions with interfering gluons.
  Version 5.1 - April 1991}'',} \textit{ Comput. Phys. Commun.} \textbf{ 67}
  (1992) 465,
  \href{http://dx.doi.org/10.1016/0010-4655(92)90055-4}{\doi{10.1016/0010-4655(92)90055-4}}.

\bibitem{Corcella:2000bw}
G.~Corcella\hrefCMSnoop {} { {et~al.}, ``{HERWIG 6: an event generator for
  hadron emission reactions with interfering gluons (including supersymmetric
  processes)}'',} \textit{ JHEP} \textbf{ 01} (2001) 010,
  \href{http://dx.doi.org/10.1088/1126-6708/2001/01/010}{\doi{10.1088/1126-6708/2001/01/010}},
  \href{http://www.arXiv.org/abs/hep-ph/0011363}{\texttt{
  arXiv:hep-ph/0011363}}.

\bibitem{Aad:2011jn}
\hrefCMSnoop {} {{ ATLAS} Collaboration, ``{Measurement of the cross-section
  for b-jets produced in association with a Z boson at $\sqrt(s)=7$ TeV with
  the ATLAS detector}'',} \textit{ Phys. Lett. B} \textbf{ 706} (2012) 295,
  \href{http://dx.doi.org/10.1016/j.physletb.2011.11.059}{\doi{10.1016/j.physletb.2011.11.059}},
  \href{http://www.arXiv.org/abs/1109.1403v2}{\texttt{ arXiv:1109.1403v2}}.

\bibitem{Aaltonen:2008mt}
\hrefCMSnoop {} {{ CDF} Collaboration, ``{Measurement of Cross Sections for b
  Jet Production in Events with a Z Boson in $p\bar{p}$ Collisions at $\sqrt{s}
  = 1.96$ TeV}'',} \textit{ Phys. Rev. D} \textbf{ 79} (2009) 052008,
  \href{http://dx.doi.org/10.1103/PhysRevD.79.052008}{\doi{10.1103/PhysRevD.79.052008}},
  \href{http://www.arXiv.org/abs/0812.4458v1}{\texttt{ arXiv:0812.4458v1}}.

\bibitem{Abazov:2010ix}
\hrefCMSnoop {} {{ D0} Collaboration, ``{A measurement of the ratio of
  inclusive cross sections $\sigma(p\bar{p}\rightarrow Z+b{\rm\, jet})/
  \sigma(p\bar{p}\rightarrow Z+{\rm jet})$ at $\sqrt{s}=1.96$ TeV}'',} \textit{
  Phys. Rev. D} \textbf{ 83} (2011) 031105,
  \href{http://dx.doi.org/10.1103/PhysRevD.83.031105}{\doi{10.1103/PhysRevD.83.031105}},
  \href{http://www.arXiv.org/abs/1010.6203v1}{\texttt{ arXiv:1010.6203v1}}.

\bibitem{CMSExperiment}
\hrefCMSnoop {} {{ CMS} Collaboration, ``The CMS experiment at the CERN LHC'',}
  \textit{ JINST} \textbf{ 03} (2008) S08004,
  \href{http://dx.doi.org/doi:10.1088/1748-0221/3/08/S08004}{\doi{doi:10.1088/1748-0221/3/08/S08004}}.

\bibitem{CMSLumi12}
\href {http://cdsweb.cern.ch/record/1434360} {{ CMS} Collaboration, ``Absolute
  Calibration of the Luminosity Measurement at {CMS}: {W}inter 2012 Update'',}
  CMS Physics Analysis Summary CMS-PAS-SMP-12-008, (2012).

\bibitem{CMS_paper_WZ}
\hrefCMSnoop {} {{ CMS} Collaboration, ``{Measurements of Inclusive W and Z
  Cross Sections in pp Collisions at $\sqrt{s}=7$ TeV}'',} \textit{ JHEP}
  \textbf{ 01} (2011) 080,
  \href{http://dx.doi.org/10.1007/JHEP01(2011)080}{\doi{10.1007/JHEP01(2011)080}},
  \href{http://www.arXiv.org/abs/1012.2466}{\texttt{ arXiv:1012.2466}}.

\bibitem{Melnikov:2006}
\hrefCMSnoop {} {{K. Melnikov and F. Petriello}, ``{Electroweak gauge boson
  production at hadron colliders through O($\alpha(s)^2$)}'',} \textit{ Phys.
  Rev. D} \textbf{ 74} (2006) 114017,
  \href{http://dx.doi.org/10.1103/PhysRevD.74.114017}{\doi{10.1103/PhysRevD.74.114017}},
  \href{http://www.arXiv.org/abs/hep-ph/0609070v1}{\texttt{
  arXiv:hep-ph/0609070v1}}.

\bibitem{TOPNLO}
\hrefCMSnoop {} {{R. Kleiss and W. J. Stirling}, ``{Top quark production at
  hadron colliders: Some useful formulae}'',} \textit{ Z. Phys. C} \textbf{ 40}
  (1988) 419,
  \href{http://dx.doi.org/10.1007/BF01548856}{\doi{10.1007/BF01548856}}.

\bibitem{EReco}
\href {http://cdsweb.cern.ch/record/1299116} {{ CMS} Collaboration, ``Electron
  Reconstruction and Identification at $\sqrt{s} = 7$ {TeV}'',} CMS Physics
  Analysis Summary CMS-PAS-EGM-10-004, (2010).

\bibitem{MuReco}
\href {http://cdsweb.cern.ch/record/1279140} {{ CMS} Collaboration,
  ``Performance of muon identification in pp collisions at $\sqrt{s}$ = 7
  {TeV}'',} CMS Physics Analysis Summary CMS-PAS-MUO-10-002, (2010).

\bibitem{CMS_PAS_PFT-09-001}
\href {http://cdsweb.cern.ch/record/1194487} {{ CMS} Collaboration,
  ``Particle--Flow Event Reconstruction in {CMS} and Performance for Jets,
  Taus, and {\MET}'',} CMS Physics Analysis Summary CMS-PAS-PFT-09-001, (2009).

\bibitem{CMS_PAS_PFT-10-001}
\href {http://cdsweb.cern.ch/record/1247373} {{ CMS} Collaboration,
  ``Commissioning of the Particle-flow Event Reconstruction with the first
  {LHC} collisions recorded in the {CMS} detector'',} CMS Physics Analysis
  Summary CMS-PAS-PFT-10-001, (2010).

\bibitem{Salam}
\hrefCMSnoop {} {M.~Cacciari, G.~Salam, and G.~Soyez, ``The anti-kt jet
  clustering algorithm'',} \textit{ JHEP} \textbf{ 04} (2008) 063,
  \href{http://dx.doi.org/10.1088/1126-6708/2008/04/063}{\doi{10.1088/1126-6708/2008/04/063}}.

\bibitem{Cacciari:2005hq}
\hrefCMSnoop {} {M.~Cacciari and G.~P. Salam, ``{Dispelling the N$^3$ myth for
  the $k_t$ jet-finder}'',} \textit{ Physics Letter B} \textbf{ 641} (2006) 57,
  \href{http://dx.doi.org/10.1016/j.physletb.2006.08.037}{\doi{10.1016/j.physletb.2006.08.037}},
  \href{http://www.arXiv.org/abs/hep-ph/0512210v2}{\texttt{
  arXiv:hep-ph/0512210v2}}.

\bibitem{FastJet}
M.~Cacciari, G.~Salam, and G.~Soyez, ``{FastJet user manual}'', 2011,
  \href{http://www.arXiv.org/abs/1111.6097}{\texttt{ arXiv:1111.6097}}.

\bibitem{Chatrchyan:2011ds}
\hrefCMSnoop {} {{CMS Collaboration}, ``{Determination of Jet Energy
  Calibration and Transverse Momentum Resolution in CMS}'',} \textit{ JINST}
  \textbf{ 6} (2011) P11002,
  \href{http://dx.doi.org/10.1088/1748-0221/6/11/P11002}{\doi{10.1088/1748-0221/6/11/P11002}},
  \href{http://www.arXiv.org/abs/1107.4277}{\texttt{ arXiv:1107.4277}}.

\bibitem{cacciari-2008-659}
\hrefCMSnoop {} {M.~Cacciari and G.~P. Salam, ``Pileup subtraction using jet
  areas'',} \textit{ Physics Letter B} \textbf{ 659} (2008) 119,
  \href{http://dx.doi.org/10.1016/j.physletb.2007.09.077}{\doi{10.1016/j.physletb.2007.09.077}},
  \href{http://www.arXiv.org/abs/0707.1378}{\texttt{ arXiv:0707.1378}}.

\bibitem{CMS_PAS_JME-10-001}
\href {http://cdsweb.cern.ch/record/1248210} {{ CMS} Collaboration, ``Jets in
  0.9 and 2.36 {TeV} pp Collisions'',} CMS Physics Analysis Summary
  CMS-PAS-JME-10-001, (2010).

\bibitem{CMS_PAS_BTV-10-001}
\href {http://cdsweb.cern.ch/record/1279144} {{ CMS} Collaboration,
  ``Commissioning of b-jet identification with pp collisions at $\sqrt{s}$ = 7
  {TeV}'',} CMS Physics Analysis Summary CMS-PAS-BTV-10-001, (2010).

\bibitem{CMS_PAS_BTV-11-002}
\href {http://cdsweb.cern.ch/record/1395489} {{ CMS} Collaboration, ``Status of
  b-tagging tools for 2011 data analysis'',} CMS Physics Analysis Summary
  CMS-PAS-BTV-11-002, (2011).

\bibitem{CMS_PAS_BTV-11-001}
\href {http://cdsweb.cern.ch/record/1366061} {{ CMS} Collaboration,
  ``Performance of b-jet identification in {CMS}'',} CMS Physics Analysis
  Summary CMS-PAS-BTV-11-001, (2011).

\bibitem{Alwall:2007fs}
J.~Alwall\hrefCMSnoop {} { {et~al.}, ``{Comparative study of various algorithms
  for the merging of parton showers and matrix elements in hadronic
  collisions}'',} \textit{ Eur. Phys. J. C} \textbf{ 53} (2008) 473,
  \href{http://dx.doi.org/10.1140/epjc/s10052-007-0490-5}{\doi{10.1140/epjc/s10052-007-0490-5}},
  \href{http://www.arXiv.org/abs/0706.2569}{\texttt{ arXiv:0706.2569}}.

\bibitem{Alwall:2008qv}
\hrefCMSnoop {} {J.~Alwall, S.~de~Visscher, and F.~Maltoni, ``{QCD radiation in
  the production of heavy colored particles at the LHC}'',} \textit{ JHEP}
  \textbf{ 02} (2009) 017,
  \href{http://dx.doi.org/10.1088/1126-6708/2009/02/017}{\doi{10.1088/1126-6708/2009/02/017}},
  \href{http://www.arXiv.org/abs/0810.5350}{\texttt{ arXiv:0810.5350}}.

\bibitem{dzeroMPI}
\hrefCMSnoop {} {{ D0} Collaboration, ``Double parton interactions in photon+3
  jet events in ppbar collisions sqrt{s}=1.96 TeV'',} \textit{ Phys. Rev. D}
  \textbf{ 81} (2010) 052012,
  \href{http://dx.doi.org/10.1103/PhysRevD.81.052012}{\doi{10.1103/PhysRevD.81.052012}},
  \href{http://www.arXiv.org/abs/0912.5104v2}{\texttt{ arXiv:0912.5104v2}}.

\bibitem{Pumplin:2002vw}
J.~Pumplin\hrefCMSnoop {} { {et~al.}, ``{New generation of parton distributions
  with uncertainties from global QCD analysis}'',} \textit{ JHEP} \textbf{ 07}
  (2002) 012,
  \href{http://dx.doi.org/10.1088/1126-6708/2002/07/012}{\doi{10.1088/1126-6708/2002/07/012}},
  \href{http://www.arXiv.org/abs/0201195v3}{\texttt{ arXiv:0201195v3}}.

\bibitem{PDF4LHC}
M.~Botje\hrefCMSnoop {} { {et~al.}, ``{The PDF4LHC Working Group Interim
  Recommendations}'',} (2011).
  \href{http://www.arXiv.org/abs/1101.0538}{\texttt{ arXiv:1101.0538}}.

\end{thebibliography}\endgroup

\cleardoublepage \appendix\section{The CMS Collaboration \label{app:collab}}\begin{sloppypar}\hyphenpenalty=5000\widowpenalty=500\clubpenalty=5000\textbf{Yerevan Physics Institute,  Yerevan,  Armenia}\\*[0pt]
S.~Chatrchyan, V.~Khachatryan, A.M.~Sirunyan, A.~Tumasyan
\vskip\cmsinstskip
\textbf{Institut f\"{u}r Hochenergiephysik der OeAW,  Wien,  Austria}\\*[0pt]
W.~Adam, T.~Bergauer, M.~Dragicevic, J.~Er\"{o}, C.~Fabjan, M.~Friedl, R.~Fr\"{u}hwirth, V.M.~Ghete, J.~Hammer\cmsAuthorMark{1}, M.~Hoch, N.~H\"{o}rmann, J.~Hrubec, M.~Jeitler, W.~Kiesenhofer, M.~Krammer, D.~Liko, I.~Mikulec, M.~Pernicka$^{\textrm{\dag}}$, B.~Rahbaran, C.~Rohringer, H.~Rohringer, R.~Sch\"{o}fbeck, J.~Strauss, A.~Taurok, F.~Teischinger, P.~Wagner, W.~Waltenberger, G.~Walzel, E.~Widl, C.-E.~Wulz
% ...
\vskip\cmsinstskip
\dag:~Deceased\\
1:~~Also at CERN, European Organization for Nuclear Research, Geneva, Switzerland\\
% ...

\end{sloppypar}
\end{document}